\documentclass[useAMS,usenatbib]{mn2e}

\usepackage{times,epsfig,natbib,amssymb,amsmath,graphics,longtable,threeparttable}

\def \lsun{\ifmmode{{\rm\ L}_\odot}\else{${\rm\ L}_\odot $}\fi}

\def \msun{\ifmmode{{\rm\ M}_\odot}\else{${\rm\ M}_\odot$}\fi}

\def \rsun{\ifmmode{{\rm\ R}_\odot}\else{${\rm\ R}_\odot$}\fi}

\newcommand{\kms}{kms$^{-1}$}                         

\def \mdot{\ifmmode{{\rm\dot{M}}}\else{${\rm\dot{M}}$}\fi}

\newcommand\am{${'}$}

\newcommand\as{${''}$}

\newcommand{\ha}{H$\alpha${}}

\newcommand{\hb}{H$\beta${}}

\title[CC SN progenitor metallicities]{Observational constraints on the progenitor 
metallicities of core-collapse supernovae\thanks{
Based on observations made with the William Herschel Telescope operated on the island
of La Palma by the Isaac Newton Group in the Spanish Observatorio del Roque de los
Muchachos of the Institute de Astrofisica de Canarias, observations
obtained with the Apache Point Observatory 3.5-meter telescope, which is owned 
and operated by the Astrophysical Research Consortium, on observations
obtained with the 
Magellan Consortium's Clay Telescope, and observations obtained with the
ESO New Technology Telescope, proposal ID: 077.C-0414}}

\author[Anderson et al.]{J. P. Anderson\thanks{E-mail:anderson@das.uchile.cl}$^{1}$, R. A. 
Covarrubias$^{2}$, P. A. James$^{3}$, M. Hamuy$^{1}$ \&\ S. M. Habergham$^{3}$\\
$^{1}$Departamento de Astronomia, Universidad de Chile, Casilla 36-D, Santiago, Chile\\
$^{2}$Anglo-Australian Observatory, P.O. Box 296, Epping, NSW 1710, Australia\\
$^{3}$Astrophysics Research Institute,
Liverpool John Moores University,
Twelve Quays House,
Egerton Wharf,
Birkenhead,
CH41 1LD,
UK}
\begin{document}

\date{}

\pagerange{\pageref{firstpage}--\pageref{lastpage}} \pubyear{2010}

\maketitle

\label{firstpage}

\begin{abstract}
We present constraints on the progenitor metallicities of core-collapse
supernovae. To date, nearly all
metallicity constraints have been inferred from indirect methods such as metallicity gradients in
host galaxies, luminosities of host galaxies, or derived global galaxy metallicities. 
Here, progenitor metallicities are derived from optical spectra taken at
the sites of nearby supernovae, from the ratio of strong emission lines found
in their host HII regions. We present results from the spectra of 74 host HII regions
and discuss the implications that these have on the nature of core-collapse
supernova progenitors.\\
Overall, while we find that the mean metallicity of type Ibc environments
is higher than that of type II events, this difference is smaller than observed in 
previous studies. There is only a 0.06 dex difference
in the mean metallicity values, at a statistical significance of $\sim$ 1.5 $\sigma$, while using 
a KS-test we find that the two metallicity distributions are marginally consistent with being
drawn from the same parent population (probability $>$10\%). This argues that progenitor metallicity is not a
dominant parameter in deciding supernovae type, with progenitor mass and/or binarity
playing a much more significant role.\\
The mean derived oxygen metallicities (12+log(O/H)) for the different supernova types, on the \cite{pet04} scale
are; 8.580 (standard error on the mean of 0.027) for the 46 type II supernovae (dominated by type II plateau);
8.616 (0.040) for 10 type Ib; and 8.626 (0.039) for 14 type Ic. Overall the types Ibc supernovae have a mean 
metallicity of 8.635 (0.026, 27 supernovae). Hence we find a slight suggestion of a metallicity sequence, in terms
of increasing progenitor metallicity going from type II through Ib and finally Ic supernovae arising from the 
highest metallicity progenitors.\\
Finally we discuss these results in the context of all current literature 
progenitor metallicity measurements, and discuss biases and selection effects that may affect
the current sample compared to overall supernova and galaxy samples.
   

\end{abstract}


\section{Introduction}
\label{intro}
Whilst supernovae (SNe) continue to be exploited to investigate the 
nature of the Universe, exact
understanding of their progenitor systems and explosion mechanisms, and how
these relate to the transient phenomena we observe remain elusive.\\
Core-collapse (CC) SNe are thought to arise from massive stars ($>$8-10 \msun)
that collapse once they have spent their nuclear fuel, the rebounding
shock-wave of this collapse powering their observed heterogenous light curve and
spectral properties. A key goal in current SN research
is to provide links between progenitor systems and observed transients.\\
In a series of papers we have attempted to constrain the nature of progenitors
through analysis of the stellar populations found in the immediate vicinity of
nearby SNe. Here we present constraints on progenitor metallicities derived
from emission line
spectra of 74 host HII regions. 

\subsection{Core-collapse supernova progenitors}
\label{ccprog}
CC SNe are classified on the basis of properties of their light
curves (e.g. \citealt{bar79}), and features found (or lacking) in their
optical spectra (see \citealt{fil97} for a review of spectral
classifications). The most common CC
class
are the SNe type II (SNII). These show hydrogen in their spectra, which is
thought to indicate that large portions of their outer
envelopes have been retained prior to explosion. SNII are further divided into a number of 
sub-types. SNIIP are the most abundant CC type and display a long plateau
($\sim$100 days) in their light curves, while other sub-types (IIL, IIb, IIn) display
differences in their light curves and spectra, that indicate differences in the
progenitor system and surrounding circumstellar material at the epoch of explosion. The
other main CC types are the SNIbc (we use this notation throughout the paper to denote the set of
events classified as Ib, Ic or Ib/c in the literature). 
These types lack
any detectable hydrogen emission (hence the classification as I rather than II), while
SNIc also lack the helium lines seen in the spectra of SNIb. The absence of
these elements is thought to imply a large amount of mass loss by the progenitors
of these SNe where they have lost much of their outer envelopes, either due
to strong stellar winds, or mass-transfer in binary systems.\\
The most direct method for determining the properties of SN progenitors is
through their identification on pre-SN images. This has had some success,
especially for the abundant SNIIP. \cite{sma09} published a compilation of
all progenitor mass estimates for SNIIP available
at the time of writing. Using a maximum
likelihood analysis of all mass estimates, plus derived upper limits, these
authors derive a mass range for SNIIP progenitors of 8.5 to 16.5\msun\ and
conclude that SNIIP arise from red supergiant progenitors. While the statistics
for the abundant SNIIP are becoming significant (when upper limits are included), only
a small number of detections or derived upper limits are available for
the other CC sub-types. Those other supernova types of most interest (with respect to the current 
study), are the SNIbc. If arising from single stars, the proposed progenitors for these
SN types are massive WR stars (e.g. \citealt{gas86}), as these stars have been stripped 
of most of their outer envelopes, hence lacking hydrogen in their pre-supernova states
as required by the spectra of SNIbc. It is also interesting to note the recent study by \cite{lel10}
comparing the spatial distributions of supernovae and WR stars within galaxies, with respect
to host galaxy $g'$-band light. These authors find that
the distributions of SNIbc explosion sites and WR stars are consistent with being drawn from the same 
parent population.
However, as summarised by \cite{sma09b}, a direct detection of
a progenitor star remains elusive, despite 10 cases of nearby SNIbc with the required pre-SN
imaging. An alternative progenitor path for producing SNIbc (i.e. progenitors with stripped
envelopes), is through lower mass binaries, where the required mass loss occurs through
mass transfer to the secondary star (e.g. \citealt{pod92}). This second progenitor
path also finds support from the relative frequency SNIbc to SNII, which, when compared 
to the initial masses of WR stars in the local group (\citealt{cro07}), is incompatible 
with all SNIbc being produced from single WR progenitor stars.
While further direct detections are shedding new light on progenitor characteristics,
the requirement of very nearby SNe means we will have to wait for definitive answers 
from these studies, especially where the less numerous SNIbc and II sub-types are concerned.
Therefore to complement these studies we must look to other avenues of research
to investigate progenitor properties. This is even more pertinent when investigating
progenitor metallicities, as currently the direct detection gives little direct evidence 
for constraining this property.\\
Host galaxy properties are commonly employed to differentiate between
different progenitor scenarios. Indeed, the first separation of SNe into CC and
thermonuclear explosions (i.e. coming from young and old stellar populations)
arose from the lack of CC SNe and presence of thermonuclear SNIa in early
type elliptical galaxies. More recently host galaxy properties have been
used to infer metallicity differences between SNII and SNIbc
(e.g. \citealt{pri08b}; see the subsequent section for additional discussion),
and have been used to derive delay time distributions for SNIa. Our
methods used here and in previous papers, are intermediate to the above two
extremes, as we use the properties of the stellar population within galaxies,
local to the SN positions, to constrain their progenitor properties.\\

\subsection{Constraints on progenitor metallicities}
\label{progZ}

The first observational evidence for metallicity differences between SNII and
SNIbc was presented by \cite{ber97} (following initial work by
\citealt{bar92}), 
who showed that SNIbc appeared to be
more centrally concentrated than SNII within host galaxies. Metallicity
gradients are found in nearly all galaxy types (see \citealt{hen99} for a
review), in the sense that higher metallicities are found in galaxy
centres. These authors therefore interpreted the increased centralisation of
SNIbc as indication that they arose from higher metallicity
progenitors. 
Interpretations of higher metallicity progenitors for SNIbc can be understood 
from a theoretical viewpoint, where higher metallicity stars have higher 
line driven winds (e.g. \citealt{pul96,kud00,mok07}), thus the progenitors lose more 
mass through these winds prior to supernova, hence removing their outer hydrogen and helium envelopes.
The radial trends above have been confirmed with increased statistics
(see \citealt{tsv04} and most recently \citealt{hak09}), furthering the
claim of higher metallicity progenitors for SNIbc. \cite{and09} (AJ09 henceforth)
investigated the radial distribution of CC SNe with
respect to the host galaxy stellar populations (previous studies focused on
distances). 
Similar trends were found with a central concentration of SNIbc
with respect to both young and old stellar populations compared with
SNII. As above this was ascribed to differences in progenitor metallicities.
However, further analysis of this sample (\citealt{hab10}, HAJ10 henceforth) has
complicated
these interpretations. Subdividing the host galaxies from AJ09 
according to the presence or absence of signs of interaction or disturbance it was
found that both SNII and SNIbc are more centrally concentrated in the
disturbed/interacting hosts. However, this effect was much more prevalent for the SNIbc
than SNII. This was interpreted as evidence for the centralisation of
star formation during galaxy interactions, and also the requirement of a top heavy IMF 
to explain the relative abundance of SNIbc over SNII. Hence, if these interpretations
are corroborated they decrease the probable importance of metallicity effects
on the distribution of the different CC SN types.\\
Other galaxy properties also correlate with
metallicity and have been used to further investigate the nature of
progenitors. The well known galaxy luminosity-metallicity relationship was used
by \cite{pra03} and updated in \cite{boi09} to investigate differences in progenitor
metallicity between SN types, both on a global host galaxy scale, and a local one using 
galaxy metallicity gradients. They found a significant difference between the overall SNII
and SNIbc populations, with apparently contradictory results regarding any differences between
SNIb and Ic; in global terms Ib were found to arise from more metal rich galaxies, while 
in terms of local estimations the opposite was true, however these differences were put
down to low number statistics.\\
All the above studies used somewhat indirect methods for inferring progenitor
metallicities. A more direct approach is to derive gas-phase metallicities
from host galaxy spectra. \cite{pri08b} presented host galaxy metallicities for
111 SNe host galaxies (all SN types), derived from SDSS spectra. Again, they found
that SNIbc appeared
to arise from more metal rich environments than SNII. However, these
metallicities were global rather than local, and therefore do not take into
account metallicity gradients found within galaxies, and how these relate
to the differing radial positions of different SN types. 
For further constraints to be made, we therefore need to 
derive metallicities \textit{at} the
positions SNe are detected. This is the analysis we present here which follows
on from a study by \cite{mod08} who derived metallicities for the immediate
environments of broad-line SNIc that were accompanied with long duration Gamma
Ray Bursts (LGRBs), and those that were not, finding that the former favoured
lower metallicity environments. Here we extend this analysis to include all the
main CC SN types, with the main aim of comparing the metallicity of SNII and
SNIbc progenitors.\\
The paper is arranged as follows; in the next section we summarise the data used for this
investigation, listing the samples involved and observations obtained. In section 3
we discuss our host HII region metallicity determinations. In section 4 we present the results
for the different supernova types. In section 5 we discuss the implications of these results, and compare them 
to those in the current 
literature. Finally in section 6 we summarise our main conclusions.

\begin{figure*}
\includegraphics[width=14cm]{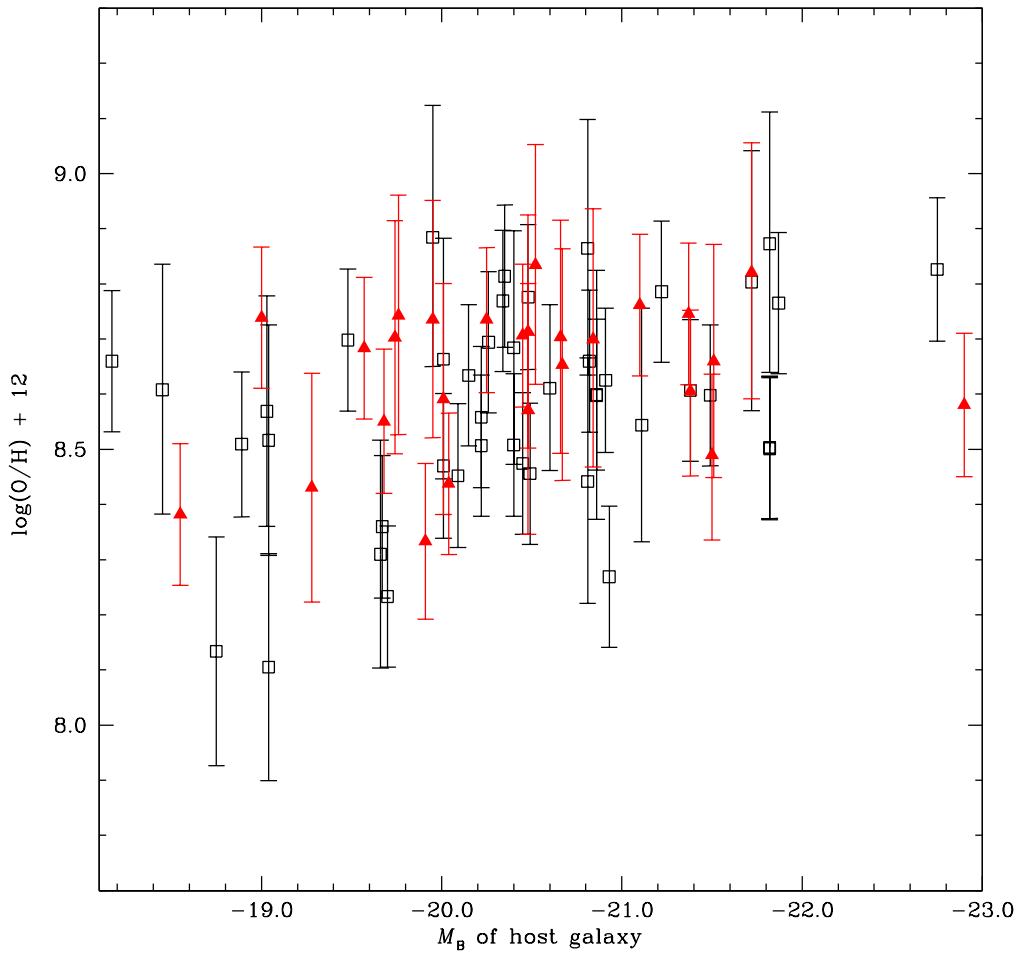}
\label{main}
\caption{Derived PP04 oxygen metallicities for CC SN host HII regions against
absolute $B$-band magnitudes of host galaxies. SNII are plotted in black squares and
the overall SNIbc population are plotted in red triangles.}
\end{figure*}

\section{Data}
\label{data}
Long slit spectra were obtained for the host HII regions of 73 CC SNe (plus 1
SN `impostor'). The
discussion of these data sets below is separated into those obtained with the ISIS
spectrograph on the William Herschel Telescope (WHT) on La Palma, in the Canary
Islands, and those obtained at various observatories, which
are taken from the thesis work of Ricardo Covarrubias (2007), the results
of which (comparing SN luminosities with environment metallicities) will be
published elsewhere (Covarrubias et al. 2010, in preparation). Many 
SNII are found far from bright HII regions (see \citealt{and08}, AJ08 henceforth), and
therefore to estimate progenitor metallicities one has to extract spectra away
from the catalogued SN positions.
Possible biases that are introduced by this procedure are discussed in 
section 5.1.\\
We note here the heterogenous nature of these data sets. The ISIS data are a
random selection taken from the SN host galaxy \ha\ imaging sample presented
in AJ09. This initial sample, in turn, was a random selection of nearby
SNe host galaxies ($<$6000 \kms) taken from the Asiago SN catalogue \citep{bar09}, 
imaged in \ha\ to investigate the association
of the different SN types to a young stellar population. We therefore assume
that both that sample, and the subset used here are a random selection of
\textit{observed} SNe in the nearby Universe. 
The Covarrubias sample consists of SNIIP that had been extensively followed 
photometrically to allow characterisation of their light curve properties and for
the use of these objects for distance measurements. These objects are, therefore, as above
a random selection of SNe discovered through different galaxy targeted searches, and therefore
are a random sample of (to date) \textit{observed} SNe in the local Universe. 
There are some cases in this second sample that are taken from galaxies further afield
than the ISIS sample, but we do not believe that these small differences will affect the 
results presented below.\\
In a perfect world one would
hope to perform such analyses as presented here on a homogeneous SN sample,
discovered through an unbiased search campaign. However, due to the proximity of galaxies
needed for this type of study (in order to detect individual HII regions, to
probe true \textit{local} metallicities), and the to date galaxy targeted
nature of nearby searches, this is not currently possible. Hence we
proceed with the current sample and discuss possible biases that may exist in 
the analysis in section 5.2.

\subsection{ISIS spectra}
\label{isis}
Long slit spectra were obtained for 50 CC SN (and 1 SN `impostor') host 
HII regions over two
observing runs with ISIS on the WHT in February 2008 and February 2009. The details of
these observations and the SN host galaxy properties are listed in Table 1. 
These data were taken with the dichroic in the default position, splitting the
incoming light into the red and blue arms at 5300 \AA. The R300B grating was
employed in the blue giving a spectral dispersion of 0.86 \AA/pixel, and the R316R in the 
red arm giving a dispersion of 0.93 \AA/pixel.
Generally data were taken with a slit width of 1\as, therefore attaining 
spectral resolutions of $\sim$4 \AA\ in each arm. However at times we observed in bad
conditions and therefore increased the slit width to 2.5\as\ to increase the target photons reaching the 
detector, decreasing the spectral resolution but we were still able to resolve \ha\ from the 
[NII] lines.\\
Most of these data were not taken at the parallactic angle, either to enable
the
host HII regions of multiple SNe to be observed, or to enable further study into
the metallicity gradients within galaxies. While some of the spectra will
therefore suffer from atmospheric differential refraction \citep{fil82}, for
metallicity determinations
we chose the line
ratio diagnostics from Pettini \&\ Pagel (2004, PP04 henceforth), who employ line ratios of emission lines
close in wavelength and hence negate this issue.
Standard reduction procedures were employed using IRAF\footnote{IRAF is 
distributed by the National Optical Astronomy Observatories,
    which are operated by the Association of Universities for Research
    in Astronomy, Inc., under cooperative agreement with the National
    Science Foundation.}
 (debiasing,
flatfielding etc) for pre-processing. Spectra were then extracted at the
closest position to the host HII region along the slit (which is $\sim$ 3.7\am\ in length). 
Extracted spectra were then wavelength
calibrated using standard arc lamps taken before or after each slit 
position, and flux calibrated using observations of spectrophotometric
standard stars taken throughout each night. Spectra were corrected
for Galactic extinction using the dust maps of \cite{sch98} and the standard 
Galactic reddening law of \cite{car89}. Where possible spectra were corrected 
for extinction internal
to the host HII region using the Balmer decrement. 
However, as stated above, because we are using line diagnostics using
the ratio of emission lines close in wavelength, the degree of these
corrections (and indeed how accurate the corrections are), should not
affect our analysis and derived metallicities, and we therefore believe them 
to be unimportant.\\
As mentioned above, many CC SNe do not fall on bright HII regions 
preventing a metallicity determination at the exact site 
of each SN detection. However, for all data we already had narrow band
\ha\ imaging, enabling us to centre the slit for spectroscopic observations on
the nearest HII region. While this may mean that for some SNe we do not
estimate the `true' progenitor stellar populations metallicity, the estimations that we present
should still be more accurate than those published previously either from
spectra integrated over whole galaxies, or those obtained from more indirect
methods as discussed earlier. 

\begin{table*}\label{isisdata} \centering
\begin{tabular}[t]{ccccccccc}
\hline
\hline
SN & Type & Host galaxy & V$_r$ (\kms) & M$_B$ & Hubble type & T-type & Slit PA & Exposure time (sec)\\
\hline
1926A & IIL & NGC 4303 & 1566 & -21.82 & SBbc & 4.2 & 159 & 3600 \\   
1941A & IIL & NGC 4559 & 816 & -21.11 & SBc & 6.2 & 132 & 3600 \\
1961U & IIL & NGC 3938 & 809 & -20.01 & Sc & 5.2 & 242 & 3600 \\
1962L & Ic & NGC 1073 & 1208 & -19.91 & SBc & 5.4 & 166 & 2700 \\ 
1964L & Ic & NGC 3938 & 809 & -20.01 & Sc & 5.2 & 5 & 3600 \\
1966J & Ib & NGC 3198 & 663 & -20.48 & SBc & 5.3 & 56 & 3600\\
1983I & Ic & NGC 4051 & 700 & -19.95 & SBbc & 4.2 & 46 & 3600\\
1984F & II & UGC 4260 & 2254 & -18.75 & Irr & 10.0 & 132 & 3600\\
1985F & Ib/c& NGC 4618 & 544 & -19.28 & SBd & 8.5 & 41 & 2700 \\
1985L & IIL & NGC 5033 & 875 & -20.86 & Sc & 5.0 & 4 & 3600 \\
1987M & Ic & NGC 2715 & 1339 & -20.67 & SBc & 5.1 & 170 & 3600\\
1990H & IIP & NGC 3294 & 1586 & -20.40 & Sc & 4.9 & 82 & 3600\\
1991N & Ic & NGC 3310 & 993 & -20.04 & SBbc & 4.0 & 330 & 2700\\
1993G & II & NGC 3690 & 3121 & -20.22 & IBm pec & 9.0 & 132 & 2700\\
1997X & Ic & NGC 4691 & 1110 & -19.57 & SB0/a & 0.0 & 93 & 2700\\
1998C & II & UGC 3825 & 8281 & -20.81 & Sbc & 4.0 & 176 & 3600\\
1998Y & II & NGC 2415 & 3784 & -21.38 & Im & 10.0 & 35 & 2700\\ 
1998cc & Ib & NGC 5172 & 4030 & -21.72 & Sbc & 4.0 & 101 & 900\\
1998T & Ib & NGC 3690 & 3121 & -20.22 & IBm pec & 9.0 & 132 & 2700\\
1999D & II & NGC 3690 & 3121 & -20.22 & IBm pec & 9.0 & 83  & 2700\\
1999bw & `imposter' & NGC 3198 & 663 & -20.48 & SBc & 5.3 & 56 & 3600\\
1999br & IIP & NGC 4900 & 960 & -19.04 & SBc & 5.0 & 107 & 4500 \\
1999ec & Ib & NGC 2207 & 2741 & -21.50 & SBbc & 4.3 & 1 & 2700 \\ 
1999ed & II & UGC 3555 & 4835 & -20.48 & Sbc & 4.0 & 122 & 1800\\
1999em & IIP & NGC 1637 & 717 & -18.45 & Sc & 5.0 & 84 & 3600\\
1999gn & IIP & NGC 4303 & 1566 & -21.82 & SBbc & 4.2 & 159 & 3600\\
2000C & Ic & NGC 2415 & 3784 & -21.38 & Im & 10.0 & 35 & 2700 \\
2001B & Ib & IC 391 & 1556 & -19.68 & Sc & 5.0 & 3 & 3600 \\
2001R & IIP & NGC 5172 & 4030 & -21.72 & Sbc & 4.0 & 101 & 900\\
2001co& Ibc & NGC 5559 & 5166 & -20.66 & SBb & 3.0 & 61 & 2700\\
2001ef & Ic & IC 381 & 2476 & -20.52 & Sbc & 3.5 & 195 & 2700 \\ 
2001ej & Ib & UGC 3829 & 4031 & -20.25 & Sb & 2.7 & 166 & 2700\\
2001gd & IIb & NGC 5033 & 875 & -20.86 & Sc & 5.0 & 4 & 3600 \\
2001is & Ib & NGC 1961 & 3934 & -22.90 & Sc & 5.0 & 157 & 3600\\
2002ce& II & NGC 2604 & 2078 & -18.89 & SBcd & 6.0 & 21 & 2700 \\
2002ji & Ib/c & NGC 3655 & 1473 & -19.76 & Sc & 5.0 & 66 & 3600\\
2002jz& Ic & UGC 2984 & 1543 & -18.55 & SBdm & 8.0 & 61 & 3600 \\
2003el& Ic & NGC 5000 & 5608 & -20.84 & SBbc & 4.0 & 94 & 3600 \\  
2004bm& Ic & NGC 3437 & 1283 & -19.74 & Sc & 5.0 & 136 & 3600 \\
2004bs & Ib & NGC 3323 & 5164 & -19.70 & SBc & 5.0 & 62 & 3600 \\
2004es& II & UGC 3825 & 8281 & -20.81 & Sbc & 4.0 & 176 & 3600 \\
2004dj & IIP & NGC 2403 & 131 & -19.66 & SBc & 5.8 & 126 & 2700\\
2004ge & Ic & UGC 3555 & 4835 & -20.48 & Sbc & 4.0 & 122 & 1800\\
2004gn& Ic & NGC 4527 & 1736 & -21.51 & SBbc & 3.8 & 64 & 3600 \\ 
2004gq& Ib & NGC 1832 & 1939 & -21.37 & SBbc & 4.0 & 56 & 2700 \\
2005V & Ib/c & NGC 2146 & 893 & -21.10 & SBab & 2.0 & 345 & 3600 \\
2005ay& IIP & NGC 3938 & 809 & -20.01 & Sc & 5.2 & 5 & 3600 \\
2005kk & II & NGC 3323 & 5164 & -19.70 & SBc & 5.0 & 62 & 3600 \\ 
2005kl& Ic & NGC 4369 & 1045 & -19.00 & Sa & 1.0 & 128 & 2700 \\
2006ov & IIP & NGC 4303 & 1566 & -21.82 & SBbc & 4.2 & 159 & 3600 \\
\hline
\end{tabular}
\caption{Table listing the SN, host galaxy properties, and observing 
information for the ISIS sub-set of the current data sample. In column
1 the SN name is listed followed by the SN type in column 2. In column 3
the SN host galaxy names are listed followed by recession velocities 
(taken from NED) then absolute $B$-band host magnitudes (taken from the LEDA database), Hubble galaxy
classifications and T-type galaxy classifications (both taken from the
Asiago catalogue). Finally in columns 8 and 9
we list the position angle of the spectrograph slit on the sky and the overall
exposure time, which was generally split into 3 separate exposures to enable
cosmic ray removal by median stacking.
}
\end{table*}

\subsection{Covarrubias sample}
\label{covar}
The second sample included in this analysis, all of host HII regions
of SNIIP (23 in total), presented in \cite{cov07} (we have also added an extra SNIIP; 2003gd,
that will appear in Covarrubias et al. 2010, in preparation), was
collected through a range of telescopes/observatories between 2004 and
2006. 
Spectroscopic data were obtained with the DIS at the APO 3.5m telescope, 
LDSS-2 and LDSS-3 on the 6.5m Clay Telescope, at LCO, and EMMI on
the ESO New Technology Telescope (NTT) at La Silla. For details of the
instrument set-ups see Covarrubias et al. (2010, in preparation). As for the ISIS data, prior \ha\ imaging
of the SN host galaxies were obtained in order to place the slit on the various
instruments on HII regions as close as possible to the position of each SN.
All these data were processed and calibrated through standard procedures as 
above, with emission line fluxes measured in preparation for
metallicity estimations. In Table 2 details of the sample
are listed. 

\begin{table*}\label{covar_data} \centering
\begin{tabular}[t]{cccccc}
\hline
\hline
SN & Host galaxy & V$_r$ (\kms) & M$_B$ & Hubble type & T-type\\
\hline
1986L & NGC 1559 & 1304 & -20.40 & SBc & 5.9\\
1990E & NGC 1035 & 1241 & -19.03 & Sc & 5.1 \\
1990K & NGC 150 & 1584 & -20.15 & SBb & 3.2\\
1991al & anon &  4572 &  & & \\
1992af & PGC 060308 & 5536 & -19.67 & Sc & 5.0 \\
1992am & PGC 3093690 & 14310 & -20.26 & Sa & 1.0 \\
1992ba & NGC 2082 & 1184 & -18.17 & SBb & 3.1 \\
1999ca & NGC 3120 & 2791 & -20.09 & Sbc & 3.7 \\
1999cr & ESO 576-034 & 6055 & -19.04 & Scd & 7.1\\
2000cb & IC 1158 & 1927 & -19.48 & Sc & 5.0  \\
2002gw & NGC 922 & 3093 & -20.93 & SBcd & 6.0 \\
2003B & NGC 1097 & 1271 & -21.22 & SBb & 3.0 \\
2003E & ESO 485-004 & 4409 & -19.70 & Sbc & 4.3\\
2003T & UGC 4864 & 8368 & -20.82 & Sab & 2.0\\
2003bj & IC 4219 & 3653 & -20.34 & SBb pec & 3.0\\
2003ci & UGC 6212 & 9105 & -21.87 & Sbc & 3.6\\
2003ef & NGC 4708 & 4166 & -20.35 & Sab pec? & 2.3\\
2003fb & UGC 11522 & 5259 & -20.91 & Sbc & 4.0\\
2003gd & NGC 628 & 657 & -20.60 & Sc & 5.0\\
2003hd & ESO 543-017 & 11842 & -21.82 & Sbc & 4.2\\
2003hk & NGC 1085 & 6789 & -21.49 & Sbc & 3.5\\
2003hl & NGC 772 &  2472 & -22.75 & Sb & 3.0\\
2003hn & NGC 1448 & 1168 & -20.49 & Sc & 5.9\\
\hline
\end{tabular}
\caption{Table listing the SNIIP and host galaxy properties
for the Covarrubias sub-sample. In column 1 the SN names are listed
followed by the host galaxy in column 2. The host galaxy properties
recession velocity, absolute $B$-band magnitude, Hubble type, and finally T-type
are given (all sources are the same as in Table 1).}
\end{table*}

\section{Host HII region metallicity determinations}
\label{anal}

\begin{table*}\label{restab} \centering
\begin{tabular}[t]{ccccc}
\hline
\hline
SN & Type & N2 & O3N2 & Distance from SN position (kpc)\\
\hline
1926A & IIL & 8.66$^{+0.21}_{-0.21}$ & 8.50$^{+0.13}_{-0.13}$ & 2.23\\
1941A & IIL & 8.54$^{+0.21}_{-0.21}$ & & SN\\
1961U & IIL & 8.47$^{+0.21}_{-0.21}$ & 8.47$^{+0.13}_{-0.13}$ & 0.75\\
1962L & Ic & 8.36$^{+0.23}_{-0.22}$ & 8.33$^{+0.14}_{-0.14}$ & 6.92\\
1964L & Ic & 8.59$^{+0.21}_{-0.21}$ & & 0.28\\
1966J & Ib & 8.57$^{+0.23}_{-0.23}$ & & 0.39\\
1983I$^*$ & Ic & 8.73$^{+0.21}_{-0.21}$ & & 2.37\\
1984F & II & 8.14$^{+0.21}_{-0.21}$ & 8.13$^{+0.13}_{-0.13}$ & 0.86\\
1985F & Ib & 8.43$^{+0.21}_{-0.21}$ & & SN\\
1985L & IIL & 8.59$^{+0.22}_{-0.22}$ & 8.60$^{+0.14}_{-0.14}$ & 2.17\\
1986L & IIP & 8.52$^{+0.21}_{-0.21}$ & 8.51$^{+0.13}_{-0.13}$ & 0.74\\
1987M & Ic & 8.65$^{+0.21}_{-0.21}$ &  & 0.03\\
1990E & IIP & 8.57$^{+0.21}_{-0.21}$ & & 2.99\\
1990H & IIP & 8.68$^{+0.21}_{-0.21}$ & & SN\\
1990K & IIP & 8.63$^{+0.21}_{-0.21}$ & 8.63$^{+0.13}_{-0.13}$ & 0.57\\
1991N & Ic & 8.49$^{+0.21}_{-0.21}$ & 8.44$^{+0.13}_{-0.13}$ & SN\\
1991al & IIP & 8.47$^{+0.21}_{-0.21}$ & 8.61$^{+0.13}_{-0.13}$ & 3.39\\
1992af & IIP & 8.37$^{+0.21}_{-0.21}$ & 8.36$^{+0.13}_{-0.13}$ & 0.38\\
1992am & IIP & 8.66$^{+0.21}_{-0.21}$ & 8.69$^{+0.13}_{-0.13}$ & 14.53\\
1992ba & IIP & 8.52$^{+0.21}_{-0.21}$ &8.66$^{+0.13}_{-0.13}$ &  0.42\\
1993G & IIL & 8.56$^{+0.21}_{-0.21}$ & 8.51$^{+0.13}_{-0.13}$ & SN\\
1997X & Ic & 8.65$^{+0.21}_{-0.21}$ & 8.68$^{+0.13}_{-0.13}$ & SN\\
1998C$^*$  & II & 8.86$^{+0.23}_{-0.23}$ & & SN\\
1998T & Ib & 8.60$^{+0.21}_{-0.21}$ & 8.53$^{+0.13}_{-0.13}$ & SN\\
1998Y & II & 8.62$^{+0.21}_{-0.21}$ & 8.61$^{+0.13}_{-0.13}$ & SN\\
1998cc & Ib & 8.82$^{+0.23}_{-0.23}$ & & 1.33\\
1999D & II & 8.75$^{+0.21}_{-0.21}$ & 8.56$^{+0.13}_{-0.13}$ & 2.64\\
1999br & IIP & 8.52$^{+0.21}_{-0.21}$ & & 0.35\\
1999bw & `imposter' & 8.50$^{+0.30}_{-0.28}$ & & 0.92\\
1999ca & IIP & 8.45$^{+0.21}_{-0.21}$ & 8.45$^{+0.13}_{-0.13}$ & 0.43\\
1999cr & IIP & 8.11$^{+0.21}_{-0.21}$ & & 6.11\\
1999ec & Ib & 8.44$^{+0.26}_{-0.25}$ & 8.45$^{+0.15}_{-0.15}$ & 0.10\\
1999ed & II & 8.79$^{+0.21}_{-0.21}$ & 8.78$^{+0.13}_{-0.13}$ & SN\\
1999em & IIP & 8.61$^{+0.23}_{-0.23}$ & & 0.23\\
1999gn$^*$ & IIP & 8.87$^{+0.24}_{-0.23}$ & & SN\\
2000C & Ic & 8.62$^{+0.21}_{-0.21}$ & 8.61$^{+0.13}_{-0.13}$ & SN\\
2000cb & IIP & 8.53$^{+0.21}_{-0.21}$ & 8.70$^{+0.13}_{-0.13}$ & 3.68\\
2001B & Ib & 8.66$^{+0.21}_{-0.21}$ & 8.55$^{+0.13}_{-0.13}$ & SN\\
2001R$^*$ & IIP & 8.80$^{+0.24}_{-0.23}$ & & 4.38\\
2001co$^*$ & Ib/c & 8.70$^{+0.21}_{-0.21}$ & & 3.01\\
2001ef$^*$ & Ic & 8.84$^{+0.23}_{-0.23}$ & & SN\\
2001ej & Ib & 8.77$^{+0.22}_{-0.21}$ & 8.74$^{+0.13}_{-0.13}$ & SN\\
2001is & Ib & 8.67$^{+0.23}_{-0.22}$ & 8.58$^{+0.13}_{-0.13}$ & 1.47\\
2001gd & IIb & 8.60$^{+0.23}_{-0.22}$ & & 0.30\\
2002ce & II & 8.49$^{+0.13}_{-0.13}$ & 8.51$^{+0.13}_{-0.13}$ & 0.07\\
2002gw & IIP & 8.24$^{+0.21}_{-0.21}$ & 8.27$^{+0.13}_{-0.13}$ & 0.81\\
2002ji & Ib/c & 8.64$^{+0.22}_{-0.22}$ & 8.74$^{+0.13}_{-0.13}$ & 0.56\\
2002jz & Ic & 8.12$^{+0.21}_{-0.21}$ & 8.38$^{+0.14}_{-0.13}$ & 0.35\\
2003B & IIP & 8.73$^{+0.21}_{-0.21}$ & 8.79$^{+0.13}_{-0.13}$ & 3.80\\
2003E & IIP & 8.14$^{+0.21}_{-0.21}$ & 8.23$^{+0.13}_{-0.13}$ & 0.28\\
2003T & IIP & 8.64$^{+0.21}_{-0.21}$ & 8.66$^{+0.13}_{-0.13}$ & 2.29\\
2003bj & IIP & 8.64$^{+0.21}_{-0.21}$ & 8.77$^{+0.13}_{-0.13}$ & 0.07\\
2003ci & IIP & 8.61$^{+0.21}_{-0.21}$ & 8.77$^{+0.13}_{-0.13}$ & 5.18\\
2003el$^*$ & Ic & 8.67$^{+0.24}_{-0.23}$ & & 0.04\\
2003ef & IIP & 8.66$^{+0.21}_{-0.21}$ & 8.81$^{+0.13}_{-0.13}$ & 1.62\\
2003ie$^*$ & II & 8.88$^{+0.24}_{-0.23}$ & & 1.18\\
2003fb & IIP & 8.58$^{+0.21}_{-0.21}$ & 8.63$^{+0.13}_{-0.13}$ & 4.52\\
2003gd & IIP & 8.54$^{+0.21}_{-0.23}$ & 8.61$^{+0.15}_{-0.15}$ & 0.20\\
\hline
\end{tabular}
\end{table*}

\setcounter{table}{2}
\begin{table*}\label{restab2} \centering
\begin{tabular}[t]{ccccc}
\hline
\hline
SN & Type & \textit{N2} & \textit{O3N2} & Distance from SN position (kpc)\\
\hline
2003hd & IIP & 8.52$^{+0.21}_{-0.21}$ & 8.50$^{+0.13}_{-0.13}$ & 1.31\\
2003hk & IIP & 8.62$^{+0.21}_{-0.21}$ & 8.60$^{+0.13}_{-0.13}$ & 13.43\\
2003hl & IIP & 8.66$^{+0.21}_{-0.21}$ & 8.87$^{+0.13}_{-0.13}$ & 1.00\\
2003hn & IIP & 8.46$^{+0.21}_{-0.21}$ & 8.46$^{+0.13}_{-0.13}$ & 3.30\\
2004bm & Ic & 8.63$^{+0.21}_{-0.21}$ & 8.70$^{+0.13}_{-0.13}$ & SN\\
2004bs & Ib & 8.62$^{+0.21}_{-0.21}$ & 8.71$^{+0.13}_{-0.13}$ & SN\\
2004dj & IIP & 8.31$^{+0.21}_{-0.21}$ & & 0.33\\
2004es & II & 8.44$^{+0.22}_{-0.22}$ & & SN\\
2004ge$^*$ & Ic & 8.71$^{+0.21}_{-0.21}$ & & SN\\
2004gq & Ib & 8.72$^{+0.21}_{-0.21}$ & 8.75$^{+0.13}_{-0.13}$ & 0.15\\
2004gn & Ic & 8.66$^{+0.21}_{-0.21}$ & & SN\\
2005V & Ib/c & 8.83$^{+0.21}_{-0.21}$ & 8.76$^{+0.13}_{-0.13}$&SN\\
2005ay$^*$ & IIP & 8.66$^{+0.23}_{-0.23}$ & &SN\\
2005kl & Ic & 8.62$^{+0.21}_{-0.21}$ & 8.74$^{+0.13}_{-0.13}$ & SN\\
2005kk & II & 8.45$^{+0.21}_{-0.21}$ & 8.48$^{+0.13}_{-0.13}$ & SN\\
2006ov & IIP & 8.66$^{+0.21}_{-0.21}$ & 8.50$^{+0.13}_{-0.13}$ &0.68\\
\hline
\end{tabular}
\caption{CC SN progenitor metallicities derived from host HII regions. In the first column the
SN name is listed followed by the SN type. Then the \textit{N2} metallicity is given together
with errors, followed by the \textit{O3N2} metallicity together with its associated errors. 
These errors are a combination of flux calibration errors, statistical errors estimated from the
variation of the continuum close to the measured line fluxes, and the 1 $\sigma$ spread in the
calibration of each line diagnostic taken from PP04. In most cases this last calibration error dominates.
In the
final column the distances between the galactocentric radial SN position and that of the HII regions
used for the metallicity derivation are listed. If these positions are the same then those cases are 
listed as `SN'. Those SNe marked with a star refer to where the metallicity determination may be somewhat
in error due to the saturation of the [NII] line (see text for further details).}
\end{table*}

SN host HII region metallicities were derived using the `empirical' methods
described in PP04. These authors derive equations relating the ratios
of [NII]6583\AA/\ha\ (\textit{N2}), and ([OIII]5007\AA/\hb)/\textit{N2} (\textit{O3N2}), to metallicity,
from calibrations using derivations of electron temperatures of HII regions within
nearby galaxies.
These diagnostics have the advantage (over those such as [NII]/[OII] and 
the commonly used $R$$_{23}$ calibrations)
of being insensitive to extinction due to
the small separation in wavelength of the emission lines used for the ratio
diagnostics. We also employ these diagnostics as they will be insensitive
to differential refraction, which is particularly pertinent for the ISIS data,
as many of these spectra were not taken at the parallactic angle, as discussed above. 
For all HII regions within our sample we therefore derive oxygen
abundances using the \textit{N2} and where possible (in some cases we did not detect
the oxygen or \hb\ emission lines), the more accurate \textit{O3N2} line ratios. These values are listed
in Table 3.
PP04 note that their calibration fit is only valid for \textit{O3N2} $<$ 1.9. There is one case
in our sample where this condition was not met (for the environment of SN 1999cr), and 
we therefore use the \textit{N2} ratio instead. We also note that above solar metallicity (12 +
log (O/H) = 8.66; \citealt{all01,asp04}) [NII] tends to saturate \citep{bal81} making \textit{N2} 
an unreliable metallicity indicator. There are 11 of these cases in the current sample,
where we only detect \ha\ and [NII], made up of 6 SNIbc and 5 SNII. While these derived metallicities 
are probably somewhat in error we are unable to derive more accurate values.
We do not think they will heavily influence the overall results presented below, and
we therefore include them in the analysis but note them with a star in Table 3 ([NII] fluxes 
are also present in the \textit{O3N2} diagnostic, however the effects of
its saturation will be diluted by the presence of other lines in the estimation, and therefore
it is thought that \textit{O3N2} will still be reliable above solar values). \\ 
In the last column of Table 3 we list the distance between the catalogued SN position
(taken from the Asiago catalogue, or derived from our own images in the case of the Covarrubias sample),
and the position of the extracted spectrum used to determine the progenitor metallicity.
This distance is calculated from the difference between the de-projected galactrocentric 
radius of the SN position, and that of the HII region where the spectrum was extracted.
If these two positions are identical (i.e. the SN fell on a bright HII region), this distance
is listed as `SN'. In section~\ref{distances} we analyse whether including those measurements
at considerable distances from the SN position may affect our results 
and conclusions. Next we present the results from the above derived CC SN progenitor metallicities.

\begin{figure*}
\includegraphics[width=14cm]{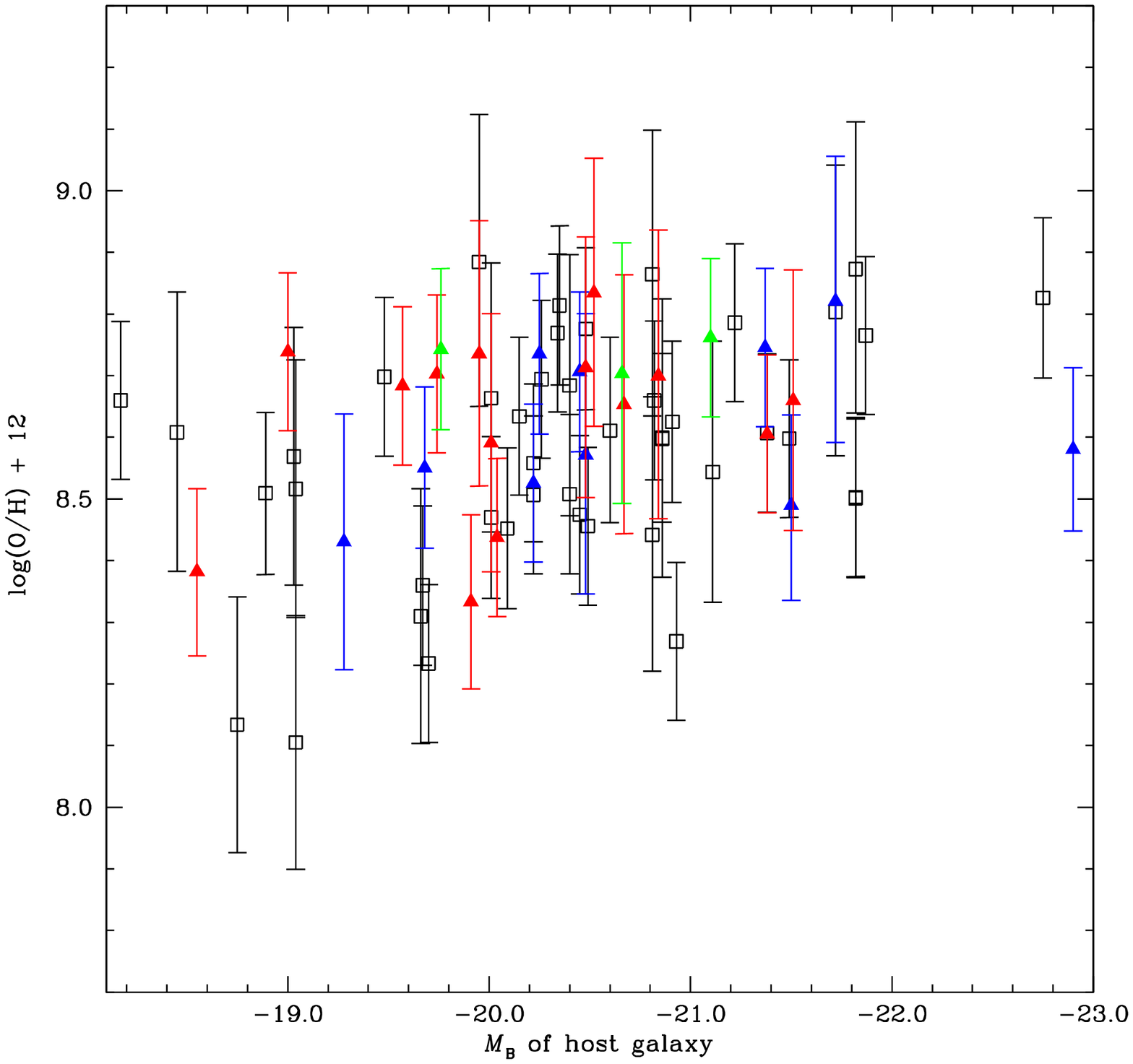}
\label{plot_all}
\caption{Derived PP04 oxygen metallicities for CC SN host HII regions against absolute
$B$-band magnitudes of host galaxies. Here we split the SNIbc into individual classes. SNII
are plotted in black squares, SNIb in blue triangles, SNIc in red triangles, and finally
those only classified as SNIb/c are plotted in green triangles.}
\end{figure*}

\section{Results}
\label{res}

\subsection{Type II supernovae}
\label{snII}
The mean oxygen abundance for the 46 SNII in the current sample, on the
PP04 scale, is 8.580 with a standard error on the mean of 0.027. This sample
spans $\sim$0.75 dex in metallicity, with SNII occurring in significant numbers at 
both high and low metallicity.
This sample is dominated by SNIIP (67\%) the mean 
metallicity for this subset being 8.587 (0.034).
In the sample of SNII there are also 4 SNIIL, which have a mean metallicity
of 8.520 (0.028), thus lower than the SNIIP, although with only 4 objects any 
difference between these types is not statistically significant. 
There is also one SNIIb (thought to be transitional objects between SNII and SNIb) in the current sample
having an \textit{N2} metallicity of 8.60$^{+0.23}_{-0.22}$. 
Finally there is also one SN `imposter' in the sample; 1998bw (which we do not include in the 
SNII sample) and this has an \textit{N2} metallicity of 8.50$^{+0.30}_{-0.28}$.

\subsection{Type Ibc supernovae}
\label{snibc}

The 10 SNIb in the sample have a mean abundance of 8.616 (0.040), while
the 14 SNIc have a mean of 8.626 (0.039).
Combining both the above samples the mean metallicity for the 27 SNIbc (adding an 
additional 3 SNe that only have Ib/c classification, note
that all these metallicities are reasonably high, hence the higher value 
of this mean than both that of the Ib and Ic samples) is 8.635 (0.026).
We therefore find that while in terms of mean metallicity values
the SNIbc seem to arise from more metal rich environments, and hence progenitors than SNII,
this difference is only marginal.\\
This is shown in Figure 1, where we plot the derived host HII region
metallicities for the SNII (black squares) and SNIbc (red triangles) against
host galaxy absolute $B$-band magnitude, which should be a rough proxy for 
global galaxy metallicity. While there is much scatter in the plot, one
can see that overall the SNIbc appear to have slightly higher derived metallicities than the SNII.
In particular the lowest 5 metallicities are derived from SNII host HII regions
(although also note that the \textit{highest} 3 metallicities are also derived from SNII
host HII regions). While there appears to be a difference in metallicity between the two
progenitor populations, this difference is not hugely significant. In terms 
of the means of the populations there is only 0.06 dex (at a level of $\sim$1.5 $\sigma$) difference
in progenitor metallicities. Using a KS test, again the difference between the 
two populations is only marginal, with $\sim$16\%\ chance that the two distributions
are drawn from the same parent population.\\
In Fig. 2 we show a similar plot but split the SNIbc into individual
classes. The SNII are shown in black squares, the SNIb in blue triangles, the SNIc
in red triangles and finally those with only Ib/c classification are shown in green triangles.
Any differences in derived metallicity between these individual classes are hard to see 
directly from this plot. However, in terms of the mean values for each class, there is
a suggestion of a metallicity sequence in terms of the SNII arising from the lowest 
metallicity progenitors, followed by the SNIb and finally the SNIc arising from the 
highest metallicity environments and hence progenitors. It is also
interesting to note that there appears to be more of a spread in metallicity for the SNIc
compared to the SNIb, something that is shown in their deviations from the mean, with the
SNIc showing a 1 $\sigma$ spread of 0.145 dex compared to 0.128 for the SNIb.\\
In terms of overall host galaxy absolute $B$-band magnitude, which should correlate
with overall galaxy metallicity, any differences between the SNIbc and SNII hosts are
even less apparent than for derived environment metallicities. The mean absolute $B$-band
magnitude for the SNII hosts is --20.42 compared to --20.43 for the SNIbc hosts. Hence in 
terms of host galaxy magnitude the two distributions are completely consistent with each other (which
is confirmed by application of the KS-test). Therefore on both a local and global scale we 
do not find conclusive evidence that the overall progenitor metallicities of SNIbc are
significantly higher than those of SNII.

\section{Discussion}
\label{diss}
The results presented above show that 
the difference in mean metallicity between the SN types is low, and statistically at best 
marginal. The simplest interpretation of this result is therefore that while
progenitor metallicity probably plays some role in deciding SN type (through increasing the 
pre-SN mass loss through line driven winds and hence stripping the progenitor
star of additional amounts of its envelope), its role is likely to be small compared to that 
played by initial progenitor mass and the role of binary companions to the
progenitor star. These statistical differences in progenitor metallicity are
again illustrated in Fig. 3. Here we plot the cumulative distribution of progenitor 
metallicities for the SNII (black), the SNIb (dashed blue), the SNIc (dashed red) and the
overall SNIbc population (green). We see that while, as discussed, the difference between the 
SNII and SNIbc distributions is not huge, there does seem to be an 
offset between the 
metallicity of the two populations. This plot also emphasises the low significance of any 
statistical difference between the SNIb and SNIc populations. It is also interesting to note that
at low metallicities ($<$$\sim$8.3) the CC SN distribution exclusively produces SNII in the
current sample.\\
The interpretation that progenitor mass plays a significantly larger role in determining
SN type is backed up by previous work using pixel statistics from host galaxy \ha\ imaging of CC SN host galaxies (AJ08), 
investigating the association of different SN types with recent star formation.
It was found, with high statistical significance that SNIbc
showed a higher degree of association to the \ha\ emission of their host galaxies than SNII.
The most logical explanation for this result is that the stellar lifetimes are significantly 
shorter for SNIbc than SNII, which implies higher mass
progenitors for the former. In a separate publication (AJ09), it was found that
SNIbc were more centrally concentrated than SNII (with respect to 
both the old stellar continuum traced by $R$-band light, and 
the young stellar host galaxy component traced by \ha\ emission), 
the initial interpretation being that this implied a significant 
progenitor metallicity difference between the two classes. Indeed, it was found that
the SNe coming from the central 20\%\ of the host galaxies' light was completely dominated by SNIbc and in
particular SNIc, implying that at high metallicities the CC SN progenitor channel highly favoured
the production of these types. However, this trend is not seen in the current work. Here we find that 
SNII are produced in significant numbers at all metallicities, with the highest three derived host HII region
metallicities being for this SN type. AJ09 also found a lack of SNIc at large galactocentric radii
implying that these SNe require high metallicity environments to achieve the required envelope
stripping to explode as SNIc. Again however, we do not see this in the current work with a number
of SNIc having relatively low derived metallicities. Indeed there are now other examples in the
literature (see further analysis and discussion in section~\ref{litZ}), of SNIc
arising from low metallicity environments. These two works therefore seem somewhat
in contradiction. However, as discussed earlier, recent further analysis of the AJ09
sample (HAJ10), splitting the sample into disturbed and undisturbed
host galaxies now indicates that this initial assumption of centralisation of
events being caused solely by metallicity differences may have been too simplistic. It was found 
that this centralisation of SNIbc events is much more pronounced in galaxies showing signs of 
disturbance or interaction. This trend is difficult to explain in terms of metallicity effects
as one would expect metallicity trends to be smaller or non-existent in galaxies undergoing some kind
of merging event where the flow of un-enriched gas to the centres of galaxies is likely to
\textit{lower} central metallicities. This work claims that the most logical explanation
is that this shows evidence for a shallower high-mass IMF in the central parts of disturbed galaxies, and hence an
increased production of SNIbc over SNII.
These results therefore very much complicate the
interpretations that can be made from differences in the positions of different SN types found within 
host galaxies. Additional work needs to be employed to differentiate between the possible effects
of changes in environment metallicity and IMF, and indeed for more reliable 
progenitor metallicity studies methods such as those employed here are needed.\\
Together with the work of \cite{mod08} (further discussed below) the most related recent study to that
presented here is that of \cite{pri08b}. As discussed earlier these authors used `global' gas-phase 
metallicities derived for SDSS galaxies host to SNe to investigate differences between 
progenitor metallicities
of different SN types. It is therefore interesting to compare the two studies. \cite{pri08b}
used a sample of 58 SNII and 19 SNIbc that had occurred in SDSS host galaxies where
spectral observations had been taken. 
These authors found a difference between the mean abundances of SNII and SNIbc of 0.12 dex; 
twice as large as the difference found here.
This may seem
slightly surprising as their analysis was of `global' metallicities
and hence does
not take into account gradients within galaxies (as we assume
our analyses should). 
Given that these are then likely to overestimate the metallicity of SNII with respect to
SNIbc (because of the differences between the radial distributions of the two types, although see section 5.4), one
would actually expect that there would be a \textit{larger} difference in mean metallicities 
in the current sample. These results are therefore slightly puzzling,
however we put these differences down to the still relatively small
sample sizes of each study and the therefore relatively low statistical significance 
($\sim$1.5 $\sigma$ here and $\sim$2 $\sigma$ in \citealt{pri08b}) of these metallicity differences.

\begin{figure}
\includegraphics[width=8cm]{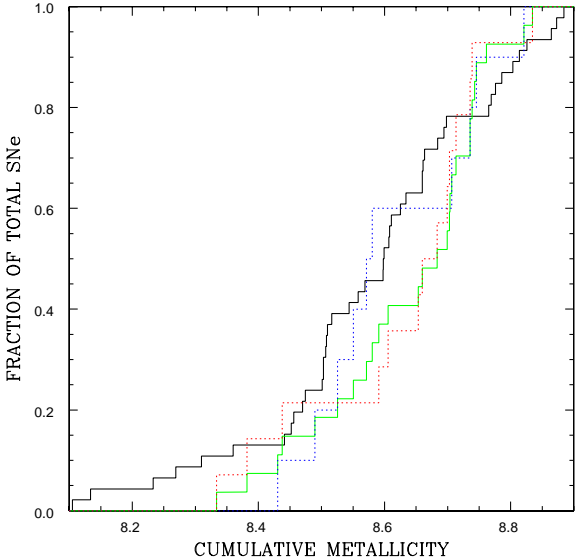}
\label{plot_cumul}
\caption{Plot showing the cumulative distribution of the metallicities of the local
gas-phase environments of the different SN types. The SNII population is shown in solid black,
the SNIb population in dashed blue, the SNIc population in dashed red, and the overall
SNIbc population in solid green.}
\end{figure}

\subsection{`Local' and `nearby' host HII regions}
\label{distances}
As we discussed earlier many SNII do not fall on bright HII regions within galaxies. While SNIbc generally show a higher 
degree of association to host galaxy \ha\ emission, even these do not always fall directly on large bright HII 
regions where one can be confident of detecting the emission lines needed for metallicity
determinations with adequate signal to noise. During the observational stage of this project
this fact was overcome by using narrow-band \ha\ imaging already in hand to locate the closest HII regions to the catalogued
SN positions, in order to place the slit of the various spectrographs employed on the desired region 
of the host galaxy. It is obvious however that this technique may adversely affect the results as metallicities may 
be determined for SN progenitors that do not reflect the `true' metallicity of the environment local
to the original SN position at the epoch of SF. Here we investigate if any such effect is at play in the above
results by splitting the sample into `local' (i.e. at or very nearby to the SN explosion coordinates) and
`nearby' (i.e. where the extraction of the spectrum used for the metallicity determinations is significantly far away from the
SN position) samples.\\
To make this separation we assume a delay time for CC SNe (time between star formation epoch and SN explosion) of 20 Myr, and a
typical progenitor velocity of 30 \kms\ (stars with velocities much higher than this are generally
considered special cases and are classified as `runaways'; \citealt{bla61}). This leads to a typical distance
traveled by a CC SN progenitor before explosion epoch of $\sim$600 pc. Therefore an accurate local metallicity 
should be considered any that is derived within this distance from the catalogued SN position. We therefore 
split the current sample into those inside and outside this limit using the values presented
in Table~\ref{restab}. We then recalculate the mean values for the SNII and SNIbc for both the `local'
and `nearby' samples. As expected the fraction of SNIbc (81\%) within this distance limit is substantially higher 
than for SNII (49\%), as SNIbc are generally found significantly closer to bright HII regions (AJ08).
The mean metallicities for SNII where the derivation was achieved within 600 pc of the SN position is 8.570, compared to
8.581 for those outside. For the SNIbc, the 5 metallicity determinations further than 600pc have exactly the same
mean metallicity for those `local' to the SN. These results mean that we can confidently include those 
metallicity determinations (38\%) that are taken away from the explosion site of the sample SNe. While these
metallicities will likely be slightly different from those at the exact site of the individual SNe the above analysis
shows that they are unlikely to bias the results in one particular direction.

\subsection{Host galaxy properties and possible biases}
\label{galaxies}
One may ask whether the sample of SNe and their host galaxies included in the current analysis are a true representation
of that occurring in nature. These SNe were found from a large number of different search programs (both amateur and
professional), all of which have different search parameters. It is therefore impossible to study any subset which
is in any way unbiased. However, we can study the host galaxy properties to understand how different the current
sample may be from that in nature. The main inadequacy of many SN searches to date is their galaxy targeted nature 
and the higher degree of time spent looking at massive nearby galaxies where one can hope to increase the likelihood
of SN detection. This in general means that smaller, lower metallicity galaxies may be underrepresented in
SN catalogues. 
However, we note that \cite{jam06} investigated this effect and found that there was no large bias
present when looking at where within the overall sample of the \ha\ Galaxy Survey \citep{jam04} SNe had been discovered, with 
respect to the star formation rate of the sample galaxies. 
Non galaxy-targeted searches are starting to find a small number of events in unclassified very small
galaxies (see e.g. \citealt{arc10} for examples of these types of SN discoveries in `rolling' searches). 
However, the contribution of these events to the overall SN rates is likely to be very small due
to the very low star formation rates that these galaxies will have. Hence this possible bias of using
SNe discovered from galaxy-targeted for overall sample studies is probably very low, if existent at all.\\
In section~\ref{data} we stated that we assumed that the current sample is a random selection of targets taken from
AJ09. While this is true, the low numbers of events may mean we are unwittingly using a subset that is different in 
host galaxy properties (and therefore may be producing different SNe), than the original parent sample. To test this we 
use the host galaxy T-types as listed in Tables 1 and 2, to calculate a mean T-type of 4.79 with a standard error on the mean of 0.23
for the current sample. 
We can then compare this to the larger sample from AJ09 which has a mean host galaxy T-type of 4.55 (0.15) and hence
within the errors the subsample presented here is consistent with the overall parent sample. We believe that this
comparison is valid because the AJ09 sample was a large fraction of the catalogued SNe occurring within $\sim$6000 \kms\
and therefore both that sample, and because of the above analysis the current sample, are a random sample of the 
\textit{observed} SNe to date. We acknowledge that there are probably certain `special' case SNe lacking from our sample, but
given the low star formation rates of their likely hosts their rates are likely low and will therefore 
not have a huge impact on studies of the type presented here.\\
Above we have discussed the results of HAJ10, where a dependence of
SN radial position (within host galaxies) on the disturbance of the host galaxy was found. This correlation has
been interpreted as evidence of a different IMF in the nuclear SF regions of these galaxies, where more SNIbc are produced at the 
expense of SNII. Therefore another source of possible bias in the current sample is whether there are more/less cases of
host galaxy disturbance than found in other samples. In  HAJ10 (where
basically the same sample as AJ09 with a small number of extra data included was used) 33\%\ of CC SNe
were found in disturbed galaxies, with the other 67\%\ found in undisturbed systems (all classifications were done visually).
In the current sample these values change to 40\%\ (disturbed) and 60\%\ (undisturbed). Hence there are more SNe found in
systems where changes in IMF may also be affecting the relative number of SNII to SNIb/c together with progenitor
metallicity. While these differences in host galaxies are only small they may affect the current results in 
reducing the impact of progenitor metallicity and increasing the impact of a changing IMF on producing the
correlations of CC SNe types on progenitor environment. Therefore we note this as a small caveat to 
the results and conclusions presented here.

\subsection{Comparison to stellar models}
\label{models}
Single star models (e.g. \citealt{heg03,eld04,mey05}) generally predict that the production rate of
SNIbc relative to SNII is correlated with progenitor metallicity. \cite{heg03} concluded that SNIbc production
required super-solar metallicities and that therefore they were not produced in significant numbers
by single star progenitors. However, their models lacked rotation, the inclusion of which
by different authors \citep{mey05} significantly lowered the required metallicity for SNIbc production. 
\cite{geo09} compared the observed ratio of SNII to SNIbc with both their single star models with rotation
and the binary models of \cite{eld08}. While this seems a worthwhile pursuit in gaining knowledge of progenitor
characteristics, the current statistics of the CC SN ratio at different metallicities means that differentiating
between stellar models (the binary models of \citealt{eld08} also predict the overall 
observed trends of increasing SNIbc to SNII ratio with metallicity) is difficult due to the
relatively small differences between the model predictions and the currently small number of SNe
with metallicity estimations (especially when binning the SNIbc/II ratio into several metallicity subsets).
We therefore refrain from comparing our results to progenitor models until we 
have obtained sufficient statistics to differentiate between different model predictions. 
It is also unclear how to directly compare 
absolute metallicities between the different observational studies (such as \citealt{pri08b}) as the use 
of different line diagnostics with significant offsets in calibrated metallicities make comparing like 
with like extremely difficult. Given the relatively large offsets between different line diagnostic methods
any comparison on absolute metallicities is perhaps premature. While we quote our absolute mean metallicities 
derived for the sample presented here, our main results do not rely on these values. The 
results presented in this paper are most significant in terms of relative \textit{differences} in metallicity.


\begin{figure*}
\includegraphics[width=14cm]{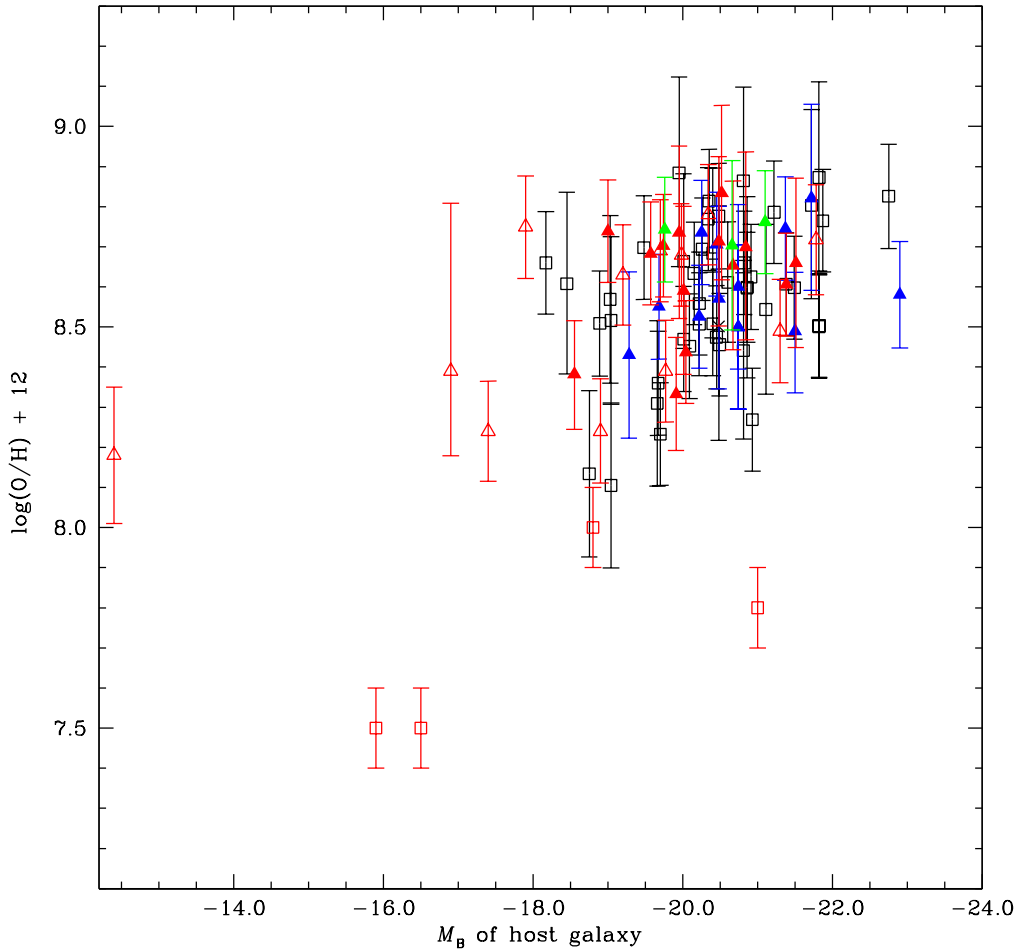}
\caption{Same as Fig. 2 except here we include all literature metallicities
as listed in Table 4. Here the SNII are shown in black squares, SNIb in blue filled triangles
SNIc in red filled triangles, SNIc-broad line in red open triangles, and SNIc-broad line with 
associated GRBs in red open squares.}
\label{lit_z}
\end{figure*}

\subsection{Analysis of Boissier \&\ Prantzos (2009)}
\label{bp09}
\cite{boi09} (henceforth BP09) recently published results on the progenitor metallicities of different
CC SN types through a study of the estimated local environment metallicity derived from global galaxy
luminosities and assumed metallicity gradients. These
authors calculated progenitor metallicities in this way for a large sample of SN host 
galaxies taken from the Asiago SN catalogue. Given that in our current sample we have 
\textit{directly} derived `local' metallicities for SN progenitors, we are
in a position to test these more indirect methods.\\
For SNe in the current sample we derived a `local' metallicity using equation (6) from
BP09 and follow these authors in taking absolute host $B$-band magnitudes from the LEDA database
and the $R$$_{25}$ radius from the Asiago catalogue. We can then compare these metallicities 
with those derived from host HII region spectra as presented here.\\
Overall we find an offset between the two methods of deriving metallicities of $\sim$0.35
dex, in terms of the BP09 metallicities being higher than those derived here. However,
this is unsurprising as their calibration uses the mass-metallicity relationship from
\cite{gar02}, where different diagnostics were used to those employed here. It is known that
the `empirical' methods we employ here give systematically lower metallicities than those 
derived from, for example photoionisation models (e.g. \citealt{kew02}). While the
origin of these differences is still debated, as long as \textit{all} metallicity
derivations are offset between different diagnostics then their use for the current 
study is justified where the main aim is to search for \textit{differences} between different SN types.
Therefore in comparing our metallicities with those determined using the BP09 technique, what is of more interest 
is the spread of these offsets. We find a 1 $\sigma$ standard deviation around this mean offset of $\sim$0.19 dex.
Calculating the mean error on our metallicity determinations listed in Table 3 (all 1 $\sigma$ errors), we
obtain a value of 0.16 dex. Hence the spread of the differences between these two methods is only slightly
larger than one would expect from the errors we estimate here from our `direct' abundance measurements.
We therefore conclude that the BP09 technique overall gives reasonable estimates for environment metallicities
which are adequate for large population scale studies. However for more precise measurements, especially
when comparing small SN samples, we encourage more observations and derived metallicities from spectra
obtained close to SN explosion positions.\\
The above analysis was performed to check the overall accuracy of the BP09 technique compared to that 
employed here, irrespective of SN type. Now we use the BP09 method to calculate mean values for the SNIbc
and SNII distributions in the current sample. Using their method we calculate a mean metallicity of 8.91 for
46 SNII, and 9.02 for 27 SNIbc, and we find a 0.1 dex metallicity
difference between the two distributions. Furthermore a using a KS-test we find only a 2.3\%\ chance that the two
distributions are drawn from the same parent population. Hence these results seem in contradiction to those
from the current study where only a marginal metallicity difference between the two SN types was found. We speculate
that this difference is related to the radial distribution of the two SN types within their host galaxies, and how
much these radial trends are related to progenitor metallicity. As discussed in detail earlier, the results of HAJ10
suggest that (at least in that sample) the overall centralisation of SNIbc in host galaxies is driven (almost entirely)
by SNe in galaxies that show signs of disturbance. Those galaxies showing signs of disturbance/evidence of interaction with a 
merging companion are likely to have much flatter (or even non-existent) metallicity gradients than their
undisturbed counteparts. Hence when the radial term in eqation (6) from BP09 is used it \textit{may} often be
overestimating the metallicity for those SNIbc found in the central parts of galaxies. 
We can investigate this assertion using the above analysis. When splitting the SNe into SNII and SNIbc we find that the offset between
our metallicity estimations and those using the BP09 method is indeed greater for SNIbc (we find an offset for SNII of 0.34 
dex and 0.38 dex for the SNIbc), and the BP09 method seems to overestimate environment metallicities for SNIb/c, at least
those in the current sample. The SNIbc used in this work
are closer to the centres of their host galaxies than the SNII (having a mean $R$/$R$$_{25}$ of 0.30 compared to 0.47 for the SNII), as 
observed in AJ09 and other previous studies,
and therefore BP09 will correct the metallicities for this trend, even though the correlation of SN type with
radial position may not be determined by metallicity. This may therefore be the source of the discrepancy. 


\subsection{Literature progenitor metallicities}
\label{litZ}
While the current study is (to our knowledge) the first to publish a large sample of 
environment metallicities for the main
CC SN types, over the last few years a number of individual progenitor metallicities have been
published by various authors, along with the GRB/SNIc sample published by \cite{mod08}. 
In this section we therefore bring all these metallicities together and discuss 
their overall implications for CC progenitor metallicities and on future SN environment studies.
These literature metallicities are listed in Table 4 together with their references. 

\begin{table*}\label{littab} \centering
\begin{tabular}[t]{cccccc}
\hline
\hline
SN/GRB & Type & Host galaxy & Galaxy \textit{M}$_B$ & Metallicity & Reference\\
\hline
1997ef & Ic BL & UGC 4107 & -19.97 & 8.69$^{+0.13}_{-0.13}$ & \cite{mod08} \\
1998ey & Ic BL & NGC 7080 & -21.78 & 8.72$^{+0.14}_{-0.14}$ & \cite{mod08}\\
1999eh & Ib & NGC 2770 & -20.74 & 8.50$^{+0.21}_{-0.21}$ & \cite{tho09}\\
GRB020819 & GRB & anon &  & 8.70$^{+0.13}_{-0.14}$ & \cite{lev10a}\\
GRB020903 & GRB & anon & -18.80 & 8.00$^{+0.10}_{-0.10}$ & \cite{mod08}\\
GRB030329/2003dh & GRB/SNIc BL & anon & -16.50 & 7.50$^{+0.10}_{-0.10}$ & \cite{mod08}\\
2003jd & Ic BL & MCG -01-59-21 & -19.77 & 8.39$^{+0.13}_{-0.13}$ & \cite{mod08}\\
GRB031203/2003lw & GRB/SNIc BL & anon & -21.00 &  7.80$^{+0.10}_{-0.10}$ & \cite{mod08}\\
2005kr & Ic BL &  J030829.66+005320.1 & -17.40 & 8.24$^{+0.13}_{-0.13}$ & \cite{mod08}\\
2005ks & Ic BL & J213756.52-000157.7 & -19.20 & 8.63$^{+0.13}_{-0.13}$ & \cite{mod08}\\
2005nb & Ic BL & UGC 7230 & -21.30 & 8.49$^{+0.13}_{-0.13}$ & \cite{mod08}\\
GRB/XRF060218/2006aj & GRB/XRF/SNIc BL & anon & -15.90 & 7.50$^{+0.10}_{-0.10}$ & \cite{mod08}\\
2006nx & Ic BL & J033330.43-004038.0 & -18.90 & 8.24$^{+0.13}_{-0.13}$ & \cite{mod08}\\
2006qk & Ic BL & J222532.38+000914.9 & -21.00 & 8.75$^{+0.13}_{-0.13}$ & \cite{mod08}\\
2007I  & Ic BL & J115913.13-013616.1 & -16.90 & 8.39$^{+0.42}_{-0.21}$ & \cite{mod08}\\
2007bg & Ic BL & anon & -12.40 & 8.18$^{+0.17}_{-0.17}$ & \cite{you10} \\
2007uy & Ib & NGC 2770 & -20.74 & 8.50$^{+0.13}_{-0.13}$ & \cite{tho09}\\
2007ru & Ic BL & UGC 12381 & -20.34 & 8.78$^{+0.13}_{-0.13}$ & \cite{sah09}\\
2008D  & Ib & NGC 2770 & -20.74 & 8.60$^{+0.21}_{-0.21}$ & \cite{tan09,tho09}\\
2009bb & Ic BL & NGC 3278 & -19.98 & 8.68$^{+0.13}_{-0.13}$ & \cite{lev10b}\\
\hline
\end{tabular}
\caption{Table listing the CC SN and GRB environment metallicities (on the PP04 scale) 
found through a search of the current literature. In the first column the SN/GRB names are listed.
This is followed by the host galaxy name and absolute $B$-band magnitude. Finally the 
environment metallicity together with its associated error are listed followed by the reference for these data in the final column.}
\end{table*}

In Fig. 4 we re-produce Fig. 2 but add those progenitor metallicities from Table 4.
The first observation that can be made from this plot is that the four lowest metallicities are those
for GRB environments. This observation, that GRBs in general favour lower metallicity environments than 
SNIc without accompanying GRB was the major conclusion from \cite{mod08}. While we do not add to the statistics
on SNIc-broad line or GRB metallicities here, the current work supports this view that \textit{generally} (although we 
note the case of GRB 020819) when compared to all SN types GRBs favour lower metallicity environments.
We also see a number of broad-line SNIc at low metallicities with four exploding in the lowest luminosity host galaxies.
While this is interesting it is probably unwise to draw firm conclusions from this observation, as many of these events
have published data \textit{because} they are interesting cases (i.e. associated with GRBs, occurring in strange environments),
and therefore it is possible that their rates when compared to other objects on the plot may be unrealistic.\\
It is clear from the above results and discussion that investigating progenitor metallicity
differences between different SN types is complicated, due to the absence of unbiased samples, and 
the uncertain nature of other factors that play a role in defining SN types from environmental conditions.
Eventually when un-targeted SN searches catch up in terms of nearby SN discoveries we will be able to reduce these biases 
and attempt to constrain the true range of environments which different SN types are found in and their relative rates.
For now we must continue to increase the sample sizes such as those presented here, in order to gain further 
knowledge of SN progenitor characteristics.

\section{Conclusions}
\label{conc}
In this paper we have presented results on the progenitor metallicities of CC SNe
derived through analysis of emission line spectra of HII regions in the 
immediate vicinity of the explosion sites. Overall we find only marginal evidence 
for a progenitor metallicity difference between SNIbc and SNII, in the sense
that the mean values are slightly higher for the SNIbc. There is a large
overlap in metallicities between the two classes with no clear transition
metallicity from SNII to SNIbc and with both SN types being found in considerable numbers
at all metallicities probed by the current study.

\section*{Acknowledgments}
We thank the referee F. Mannucci for his constructive comments and suggestions on the paper.
M.H. acknowledges support from Fondecyt through grant 1060808, Centro de 
Astrof\'\i sica FONDAP 15010003, Center of Excellence in Astrophysics and 
Associated Technologies (PFB 06). J.A. and M.H. acknowledge support from 
the Millennium Center for Supernova Science through grant P06-045-F.
This research
has made use of the NASA/IPAC Extragalactic Database (NED) which is operated by the Jet Propulsion Laboratory, California
Institute of Technology, under contract with the National Aeronautics and Space Administration.
We also acknowledge the use of the HyperLeda database (http://leda.univ-lyon1.fr).

\bibliographystyle{mn2e}

\bibliography{Reference}

\begin{thebibliography}{}

\bibitem[\protect\citeauthoryear{{Allende Prieto}, {Lambert} \&
  {Asplund}}{{Allende Prieto} et~al.}{2001}]{all01}
{Allende Prieto} C.,  {Lambert} D.~L.,    {Asplund} M.,  2001, \apjl, 556, L63

\bibitem[\protect\citeauthoryear{{Anderson} \& {James}}{{Anderson} \&
  {James}}{2008}]{and08}
{Anderson} J.~P.,  {James} P.~A.,  2008, \mnras, 390, 1527

\bibitem[\protect\citeauthoryear{{Anderson} \& {James}}{{Anderson} \&
  {James}}{2009}]{and09}
{Anderson} J.~P.,  {James} P.~A.,  2009, \mnras, 399, 559

\bibitem[\protect\citeauthoryear{{Arcavi} et~al.,}{{Arcavi}
  et~al.}{2010}]{arc10}
{Arcavi} I.,  et~al., 2010, ArXiv e-prints

\bibitem[\protect\citeauthoryear{{Asplund}, {Grevesse}, {Sauval}, {Allende
  Prieto} \& {Kiselman}}{{Asplund} et~al.}{2004}]{asp04}
{Asplund} M.,  {Grevesse} N.,  {Sauval} A.~J.,  {Allende Prieto} C.,
  {Kiselman} D.,  2004, \aap, 417, 751

\bibitem[\protect\citeauthoryear{{Baldwin}, {Phillips} \&
  {Terlevich}}{{Baldwin} et~al.}{1981}]{bal81}
{Baldwin} J.~A.,  {Phillips} M.~M.,    {Terlevich} R.,  1981, \pasp, 93, 5

\bibitem[\protect\citeauthoryear{{Barbon}, {Buondi}, {Cappellaro} \&
  {Turatto}}{{Barbon} et~al.}{2009}]{bar09}
{Barbon} R.,  {Buondi} V.,  {Cappellaro} E.,    {Turatto} M.,  2009, VizieR
  Online Data Catalog, 1, 2024

\bibitem[\protect\citeauthoryear{{Barbon}, {Ciatti} \& {Rosino}}{{Barbon}
  et~al.}{1979}]{bar79}
{Barbon} R.,  {Ciatti} F.,    {Rosino} L.,  1979, \aap, 72, 287

\bibitem[\protect\citeauthoryear{{Bartunov}, {Makarova} \&
  {Tsvetkov}}{{Bartunov} et~al.}{1992}]{bar92}
{Bartunov} O.~S.,  {Makarova} I.~N.,    {Tsvetkov} D.~I.,  1992, \aap, 264, 428

\bibitem[\protect\citeauthoryear{{Blaauw}}{{Blaauw}}{1961}]{bla61}
{Blaauw} A.,  1961, \bain, 15, 265

\bibitem[\protect\citeauthoryear{{Boissier} \& {Prantzos}}{{Boissier} \&
  {Prantzos}}{2009}]{boi09}
{Boissier} S.,  {Prantzos} N.,  2009, \aap, 503, 137

\bibitem[\protect\citeauthoryear{{Cardelli}, {Clayton} \& {Mathis}}{{Cardelli}
  et~al.}{1989}]{car89}
{Cardelli} J.~A.,  {Clayton} G.~C.,    {Mathis} J.~S.,  1989, \apj, 345, 245

\bibitem[\protect\citeauthoryear{{Covarrubias}}{{Covarrubias}}{2007}]{cov07}
{Covarrubias} R.~A.,  2007, PhD thesis, University of Washington

\bibitem[\protect\citeauthoryear{{Crowther}}{{Crowther}}{2007}]{cro07}
{Crowther} P.~A.,  2007, \araa, 45, 177

\bibitem[\protect\citeauthoryear{{Eldridge}, {Izzard} \& {Tout}}{{Eldridge}
  et~al.}{2008}]{eld08}
{Eldridge} J.~J.,  {Izzard} R.~G.,    {Tout} C.~A.,  2008, \mnras, 384, 1109

\bibitem[\protect\citeauthoryear{{Eldridge} \& {Tout}}{{Eldridge} \&
  {Tout}}{2004}]{eld04}
{Eldridge} J.~J.,  {Tout} C.~A.,  2004, \mnras, 353, 87

\bibitem[\protect\citeauthoryear{{Filippenko}}{{Filippenko}}{1982}]{fil82}
{Filippenko} A.~V.,  1982, \pasp, 94, 715

\bibitem[\protect\citeauthoryear{{Filippenko}}{{Filippenko}}{1997}]{fil97}
{Filippenko} A.~V.,  1997, \araa, 35, 309

\bibitem[\protect\citeauthoryear{{Garnett}}{{Garnett}}{2002}]{gar02}
{Garnett} D.~R.,  2002, \apj, 581, 1019

\bibitem[\protect\citeauthoryear{{Gaskell}, {Cappellaro}, {Dinerstein},
  {Garnett}, {Harkness} \& {Wheeler}}{{Gaskell} et~al.}{1986}]{gas86}
{Gaskell} C.~M.,  {Cappellaro} E.,  {Dinerstein} H.~L.,  {Garnett} D.~R.,
  {Harkness} R.~P.,    {Wheeler} J.~C.,  1986, \apjl, 306, L77

\bibitem[\protect\citeauthoryear{{Georgy}, {Meynet}, {Walder}, {Folini} \&
  {Maeder}}{{Georgy} et~al.}{2009}]{geo09}
{Georgy} C.,  {Meynet} G.,  {Walder} R.,  {Folini} D.,    {Maeder} A.,  2009,
  \aap, 502, 611

\bibitem[\protect\citeauthoryear{{Habergham}, {Anderson} \&
  {James}}{{Habergham} et~al.}{2010}]{hab10}
{Habergham} S.~M.,  {Anderson} J.~P.,    {James} P.~A.,  2010, ArXiv e-prints

\bibitem[\protect\citeauthoryear{{Hakobyan}, {Mamon}, {Petrosian}, {Kunth} \&
  {Turatto}}{{Hakobyan} et~al.}{2009}]{hak09}
{Hakobyan} A.~A.,  {Mamon} G.~A.,  {Petrosian} A.~R.,  {Kunth} D.,    {Turatto}
  M.,  2009, \aap, 508, 1259

\bibitem[\protect\citeauthoryear{{Heger}, {Fryer}, {Woosley}, {Langer} \&
  {Hartmann}}{{Heger} et~al.}{2003}]{heg03}
{Heger} A.,  {Fryer} C.~L.,  {Woosley} S.~E.,  {Langer} N.,    {Hartmann}
  D.~H.,  2003, \apj, 591, 288

\bibitem[\protect\citeauthoryear{{Henry} \& {Worthey}}{{Henry} \&
  {Worthey}}{1999}]{hen99}
{Henry} R.~B.~C.,  {Worthey} G.,  1999, \pasp, 111, 919

\bibitem[\protect\citeauthoryear{{James} \& {Anderson}}{{James} \&
  {Anderson}}{2006}]{jam06}
{James} P.~A.,  {Anderson} J.~P.,  2006, \aap, 453, 57

\bibitem[\protect\citeauthoryear{{James} et~al.,}{{James}
  et~al.}{2004}]{jam04}
{James} P.~A.,  et~al., 2004, \aap, 414, 23

\bibitem[\protect\citeauthoryear{{Kewley} \& {Dopita}}{{Kewley} \&
  {Dopita}}{2002}]{kew02}
{Kewley} L.~J.,  {Dopita} M.~A.,  2002, \apjs, 142, 35

\bibitem[\protect\citeauthoryear{{Kudritzki} \& {Puls}}{{Kudritzki} \&
  {Puls}}{2000}]{kud00}
{Kudritzki} R.-P.,  {Puls} J.,  2000, \araa, 38, 613

\bibitem[\protect\citeauthoryear{{Leloudas}, {Sollerman}, {Levan}, {Fynbo},
  {Malesani} \& {Maund}}{{Leloudas} et~al.}{2010}]{lel10}
{Leloudas} G.,  {Sollerman} J.,  {Levan} A.~J.,  {Fynbo} J.~P.~U.,  {Malesani}
  D.,    {Maund} J.~R.,  2010, ArXiv e-prints

\bibitem[\protect\citeauthoryear{{Levesque} et~al.,}{{Levesque}
  et~al.}{2010}]{lev10b}
{Levesque} E.~M.,  et~al., 2010, \apjl, 709, L26

\bibitem[\protect\citeauthoryear{{Levesque}, {Kewley}, {Graham} \&
  {Fruchter}}{{Levesque} et~al.}{2010}]{lev10a}
{Levesque} E.~M.,  {Kewley} L.~J.,  {Graham} J.~F.,    {Fruchter} A.~S.,  2010,
  \apjl, 712, L26

\bibitem[\protect\citeauthoryear{{Meynet} \& {Maeder}}{{Meynet} \&
  {Maeder}}{2005}]{mey05}
{Meynet} G.,  {Maeder} A.,  2005, \aap, 429, 581

\bibitem[\protect\citeauthoryear{{Modjaz}, {Kewley}, {Kirshner}, {Stanek},
  {Challis}, {Garnavich}, {Greene}, {Kelly} \& {Prieto}}{{Modjaz}
  et~al.}{2008}]{mod08}
{Modjaz} M.,  {Kewley} L.,  {Kirshner} R.~P.,  {Stanek} K.~Z.,  {Challis} P.,
  {Garnavich} P.~M.,  {Greene} J.~E.,  {Kelly} P.~L.,    {Prieto} J.~L.,  2008,
  \aj, 135, 1136

\bibitem[\protect\citeauthoryear{{Mokiem} et~al.,}{{Mokiem}
  et~al.}{2007}]{mok07}
{Mokiem} M.~R.,  et~al., 2007, \aap, 473, 603

\bibitem[\protect\citeauthoryear{{Pettini} \& {Pagel}}{{Pettini} \&
  {Pagel}}{2004}]{pet04}
{Pettini} M.,  {Pagel} B.~E.~J.,  2004, \mnras, 348, L59

\bibitem[\protect\citeauthoryear{{Podsiadlowski}, {Joss} \&
  {Hsu}}{{Podsiadlowski} et~al.}{1992}]{pod92}
{Podsiadlowski} P.,  {Joss} P.~C.,    {Hsu} J.~J.~L.,  1992, \apj, 391, 246

\bibitem[\protect\citeauthoryear{{Prantzos} \& {Boissier}}{{Prantzos} \&
  {Boissier}}{2003}]{pra03}
{Prantzos} N.,  {Boissier} S.,  2003, \aap, 406, 259

\bibitem[\protect\citeauthoryear{{Prieto}, {Stanek} \& {Beacom}}{{Prieto}
  et~al.}{2008}]{pri08b}
{Prieto} J.~L.,  {Stanek} K.~Z.,    {Beacom} J.~F.,  2008, \apj, 673, 999

\bibitem[\protect\citeauthoryear{{Puls} et~al.,}{{Puls}  et~al.}{1996}]{pul96}
{Puls} J.,  et~al., 1996, \aap, 305, 171

\bibitem[\protect\citeauthoryear{{Sahu}, {Tanaka}, {Anupama}, {Gurugubelli} \&
  {Nomoto}}{{Sahu} et~al.}{2009}]{sah09}
{Sahu} D.~K.,  {Tanaka} M.,  {Anupama} G.~C.,  {Gurugubelli} U.~K.,    {Nomoto}
  K.,  2009, \apj, 697, 676

\bibitem[\protect\citeauthoryear{{Schlegel}, {Finkbeiner} \&
  {Davis}}{{Schlegel} et~al.}{1998}]{sch98}
{Schlegel} D.~J.,  {Finkbeiner} D.~P.,    {Davis} M.,  1998, \apj, 500, 525

\bibitem[\protect\citeauthoryear{{Smartt}}{{Smartt}}{2009}]{sma09b}
{Smartt} S.~J.,  2009, \araa, 47, 63

\bibitem[\protect\citeauthoryear{{Smartt}, {Eldridge}, {Crockett} \&
  {Maund}}{{Smartt} et~al.}{2009}]{sma09}
{Smartt} S.~J.,  {Eldridge} J.~J.,  {Crockett} R.~M.,    {Maund} J.~R.,  2009,
  \mnras, 395, 1409

\bibitem[\protect\citeauthoryear{{Tanaka} et~al.,}{{Tanaka}
  et~al.}{2009}]{tan09}
{Tanaka} M.,  et~al., 2009, \apj, 700, 1680

\bibitem[\protect\citeauthoryear{{Th{\"o}ne}, {Micha{\l}owski}, {Leloudas},
  {Cox}, {Fynbo}, {Sollerman}, {Hjorth} \& {Vreeswijk}}{{Th{\"o}ne}
  et~al.}{2009}]{tho09}
{Th{\"o}ne} C.~C.,  {Micha{\l}owski} M.~J.,  {Leloudas} G.,  {Cox} N.~L.~J.,
  {Fynbo} J.~P.~U.,  {Sollerman} J.,  {Hjorth} J.,    {Vreeswijk} P.~M.,  2009,
  \apj, 698, 1307

\bibitem[\protect\citeauthoryear{{Tsvetkov}, {Pavlyuk} \&
  {Bartunov}}{{Tsvetkov} et~al.}{2004}]{tsv04}
{Tsvetkov} D.~Y.,  {Pavlyuk} N.~N.,    {Bartunov} O.~S.,  2004, Astronomy
  Letters, 30, 729

\bibitem[\protect\citeauthoryear{{van den Bergh}}{{van den
  Bergh}}{1997}]{ber97}
{van den Bergh} S.,  1997, \aj, 113, 197

\bibitem[\protect\citeauthoryear{{Young} et~al.,}{{Young}
  et~al.}{2010}]{you10}
{Young} D.~R.,  et~al., 2010, \aap, 512, A70+

\end{thebibliography}

\appendix

\label{lastpage}

\end{document}